\documentstyle[eqsecnum,prd,aps,epsfig]{revtex}
\newcommand{\dfr}{d\raise0.3ex\hbox{\kern-0.6ex\char"013 }} 
\newcommand{\ddfr}{{ }^{-}\!\!\!\!\!\dfr} 
\begin{document}
%
%
\def\ov{\over}
\def\l{\left}
\def\r{\right}
\def\be{\begin{equation}}
\def\ee{\end{equation}}
\draft
\title{A scale-relativistic derivation of the Dirac Equation}
\author{Marie-No\"elle C\'el\'erier and Laurent Nottale}
\address{LUTH, CNRS, Observatoire de Paris-Meudon, 
5 place Jules Janssen, 92195 Meudon Cedex, France}

\date{\today}
\maketitle

\begin{abstract}
The application of the theory of scale relativity to microphysics aims 
at recovering quantum mechanics as a new non-classical mechanics on a 
non-derivable space-time. This program was already achieved as regards 
the Schr\"odinger and Klein Gordon equations, which have been 
derived in terms of geodesic equations in this framework: namely, 
they have been written according to a generalized equivalence/strong 
covariance principle in the form of free motion equations 
$D^2x/ds^2=0$, where $D/ds$ are covariant derivatives built 
from the description of the fractal/non-derivable geometry. Following 
the same line of thought and using the mathematical tool of Hamilton's 
bi-quaternions, we propose here a derivation of the Dirac equation 
also from a geodesic equation (while it is still merely postulated in 
standard quantum physics). The complex nature of the wave function in the 
Schr\"odinger and Klein-Gordon equations was deduced from the necessity 
to introduce, because of the non-derivability, a discrete symmetry 
breaking on the proper time differential element. By extension, the 
bi-quaternionic nature of the Dirac bi-spinors arises here from further 
discrete symmetry breakings on the space-time variables, which also 
proceed from non-derivability.

\end{abstract}


\section{Introduction}
\label{s:intro}

The theory of scale relativity generalizes to 
scale transformations the principle of relativity, which has been applied 
by Einstein to motion laws. It is based on the giving up of the assumption 
of space-time coordinate differentiability, which is usually retained as an 
implicit hypothesis in current physics. Even though this hypothesis can be 
considered as roughly valid in the classical domain, it is clearly broken 
by the quantum mechanical behavior. It has indeed been pointed out by 
Feynman (see, e.g., Ref. \cite{FH65}) that the typical paths of quantum 
mechanics are continuous but non-differentiable. 
In the present paper, we apply the scale-relativistic approach to the 
understanding of the nature of the bi-spinors and of the Dirac 
equation in a space-time representation. \\

One of the fundamental reasons for jumping to a scale-relativistic description 
of nature is to increase the generality of the first principles 
retained. Einstein's principle of relativity relies on the assumption that 
coordinates are a priori differentiable. It is therefore not fully general, 
since it applies to coordinate 
transformations that are continuous and, at least two times, differentiable. 
The aim of the theory of scale relativity is to bring to light laws and 
structures that would be the manifestation of more general 
transformations, namely, continuous ones, either differentiable or not. The 
standard ``general relativistic'' theory will be automatically recovered as 
the special differentiable case, while new effects would be expected from the 
non-differentiable part. \\

Giving up the assumption of differentiability has important physical 
consequences: one can show \cite{LN93,LN94A} that spaces of topological 
dimension $D_T$, which are continuous but non-differentiable, are 
characterized by a $D_T$ measure which becomes explicitly dependent on 
the resolution (i.e., the observation scale) $\epsilon$ at which it is 
considered and tends to infinity 
when the resolution interval $\epsilon$ tends to zero. Therefore, 
a non-differentiable space-time continuum is necessarily fractal, in the 
general meaning given to this word by Mandelbrot \cite{BM82}, who thus 
means, not only self-similar, but explicitly scale-dependent in the more 
general way. This statement 
naturally leads to the proposal of a geometric tool adapted to construct 
a theory based on such premises, namely, a fractal space-time (see, e.g., 
Refs. \cite{LN93,GO83,NS84,LN89,EN92,BC00}). 
Such an explicit and fundamental dependence of physical laws on the 
internal ``resolution'' scales leads to constrain the new scale laws 
(that need to be constructed) by a ``principle of scale relativity'' 
\cite{LN93}. This principle extends to scale transformations (i.e., 
contractions and dilations of resolutions) the Galileo-Einstein 
principle of relativity, that was up to now applied only to 
displacements and motion. 
This allows to include the resolutions in the definition 
of the state of the reference system and to require scale covariance of 
the equations of physics under scale transformations. A main consequence 
is that the geodesics of such a non-differentiable space-time are 
themselves fractal and in infinite number between any two points. This 
strongly suggests that the description could borrow some tools from the 
mathematical formalism of statistics, i.e., will 
provide a probabilistic interpretation, therefore naturally extending the 
realm of quantum behaviour to a larger spectrum of available scales. 
The domains of application are typically the asymptotic scale ranges of 
physics: (i) {\it Microphysics}: small length-scales and time-scales. (ii) 
{\it Cosmology}: large length-scales. (iii) {\it Complex chaotic systems}: 
long time-scales. \\

In earlier works \cite{LN93,LN96} 
the Schr\"odinger equation and the (free) Klein-Gordon equation have 
been established, from the first principles of this theory, as geodesic 
equations on a fractal space/space-time. The third main evolution equation 
of standard quantum physics, the (free) Dirac equation is here derived, 
following the same trend of thought. \\

As in these earlier works, the general method followed here, in order to reach 
a genuine demonstration of the Dirac equation from first principles, 
involves the following steps: 
(i) introduce a scale-space of internal resolutions, aimed at 
describing the underlying fractal structures of geodesic trajectories 
(that are identified with the ``particles'' in this approach); 
(ii) perform this description by writing scale differential equations 
coming under the principle of relativity, then solving them; 
(iii) express the effects induced on the laws of motion (space-time 
displacement) by the internal fractal structures of the underlying 
scale-space in terms of a covariant derivative $\ddfr /ds$; when  
applied on the classical position variable, this covariant derivative changes 
the nature of the classical velocity, which becomes complex (Schr\"odinger 
and Klein-Gordon equations) , quaternionic (Pauli equation), then 
bi-quaternionic (Dirac equation) at various depth levels of the theory; 
(iv) use the principle of relativity in the form of a generalized 
equivalence/strong covariance principle in order to write 
the motion equation as a geodesic equation, i.e. as a free motion 
equation of the form $\ddfr^2 x^{\mu}/ds^2=0$; 
(v) expand the covariant derivative and make a change of variable in 
order to introduce the quantum tool (the wave function is defined as a 
re-expression of the action), then integrate and obtain the fundamental 
equations of quantum mechanics. \\

In Sec.~\ref{s:schro}, we actualize the former works dealing with this 
issue, proposing a more accurate and profound interpretation of the nature 
of the transition from the non-differentiable (fractal scales) to the 
differentiable (classical scales) domain, and carefully justifying the 
different key choices made at each main step of the reasoning. Then 
we derive, in Sec.~\ref{s:kgeq}, the bi-quaternionic form of 
the Klein-Gordon equation, from which the Dirac equation naturally 
proceeds, as shown in Sec.~\ref{s:dieq}. Section.~\ref{s:conc} is devoted 
to the conclusion.

\section{The fractal to classical scale transition revisited}
\label{s:schro}

\subsection{Scale invariance and Galilean scale relativity}
\label{ss:galsr}

A power law scale dependence, frequently encountered in natural 
systems, is described geometrically in terms of fractals \cite{BM82,BM75}, 
and algebrically in terms of the renormalization group \cite{KW75,KW79}. 
As we show below, such simple scale-invariant laws can be identified 
with the ``Galilean" version of scale-relativistic laws. \\

Consider a non-differentiable (fractal) curvilinear coordinate 
${\cal L}(x,\epsilon)$, 
that depends on some space-time variables $x$ and on the resolution 
$\epsilon$. Such a coordinate generalizes to non-differentiable and fractal 
space-times the concept of curvilinear coordinates introduced for curved 
Riemannian space-times in Einstein's general relativity \cite{LN93}. Rather 
than considering only the strict non-differentiable mathematical object 
${\cal L}(x)$, we are interested in its different versions at various 
resolutions $\epsilon$. 
Such a point of view is particularly well-adapted to applications in physics, 
as any real measurement is always performed at a finite resolution. In this 
framework, ${\cal L}(x)$ becomes the limit when $\epsilon \rightarrow 0$ 
of the family of functions ${\cal L}(x,\epsilon)$. But while ${\cal L}(x,0)$ 
is non-differentiable, ${\cal L}(x,\epsilon)$, which we call a fractal 
function \cite{BC00,JC01}, is differentiable for all $\epsilon \neq 0$. The 
physics of a given process will therefore be completely described if we can 
know ${\cal L}(x,\epsilon)$ for all values of $\epsilon$. 
${\cal L}(x,\epsilon)$, being differentiable when $\epsilon \neq 0$, can 
be the solution to differential equations involving the derivatives of 
${\cal L}$ with respect to both $x$ and $\epsilon$. \\

Let us now apply an infinitesimal dilation $\epsilon \rightarrow 
\epsilon '=\epsilon(1+d\rho)$ to the resolution. Being, at this stage, 
interested in pure scale laws, we omit the $x$ dependence in order to 
simplify the notation and obtain, at first order,
\be
{\cal L}(\epsilon ')={\cal L}(\epsilon +\epsilon d\rho)={\cal L}(\epsilon)+
{{\partial {\cal L}(\epsilon)}\over {\partial \epsilon}} \epsilon d\rho=
(1+\tilde{D} d\rho){\cal L}(\epsilon),
\label{eq.1}
\ee
where $\tilde{D}$ is, by definition, the dilation operator. The identification 
of the two last members of this equation yields 
\be
\tilde{D}=\epsilon{\partial \over {\partial \epsilon}}={\partial \over 
{\partial 
\ln \epsilon}} \; .
\label{eq.2}
\ee

This well-known form of the infinitesimal dilation operator shows that the 
``natural'' variable for the resolution is $\ln \epsilon$, and that the 
expected new differential equations will involve quantities such as 
$\partial {\cal L}(x,\epsilon)/\partial \ln \epsilon$. An example of equations 
describing scale dependence is given by the renormalization group 
equations, as first proposed by Wilson \cite{KW75,KW79}. \\

We limit the present approach to the consideration of the simplest 
form that can be exhibited by this class of equations, leaving to future 
works the study of more complete cases. Let us built such an equation by 
stating that the variation of ${\cal L}$ under 
an infinitesimal scale transformation $d\ln\epsilon$ depends only on 
${\cal L}$ itself; namely, ${\cal L}$ determines the whole physical behavior, 
including the behavior under scale transformations. We thus write
\be
{\partial {\cal L}(x,\epsilon)\over {\partial \ln \epsilon}}=\beta(\cal L).
\label{eq.3}
\ee

As we are interested in the simplest form for such an equation, we expand 
$\beta(\cal L)$ in powers of $\cal L$. This can always be done since 
$\cal L$ may be renormalized, dividing it by its largest value, 
in such a way that the new variable  $\cal L$ remains $\ll 1$ in its 
variation domain. We obtain, to the first order, the linear equation
\be
{\partial {\cal L}(x,\epsilon)\over {\partial \ln \epsilon}}=a+b\cal L \; ,
\label{eq.4}
\ee
of which the solution is
\be
{\cal L}(x,\epsilon) = 
{\cal L}_0(x)\; \left[1+\zeta (x) \left( \frac{\lambda}{\epsilon} 
\right)^{-b}\right],
\label{eq.5}
\ee
where $\lambda ^{-b} \zeta(x)$ is an integration constant and 
${\cal L}_0=-a/b$. \\

For $b<0$, Eq.~(\ref{eq.5}) gives a fractal scale-invariant behavior at 
small scales ($\epsilon \ll \lambda$), with a fractal dimension 
$D=D_T-b$. For a curvilinear coordinate ${\cal L}$, the topological dimension 
$D_T$ is equal to $1$. The scale dimension, $\delta = D - D_T$, is defined, 
following Mandelbrot \cite{BM82,BM75}, as 
\be
\delta = \frac{d \ln {\cal L}}{d \ln ({\lambda/ \epsilon})} \; .
\label{eq.6}
\ee

When $\delta$ is constant, one obtains an asymptotic power law resolution 
dependence 
\be
{\cal L}(x,\epsilon) = {\cal L}_0(x) \left(\lambda\over\epsilon\right)^\delta.
\label{eq.7}
\ee

Let us now check that for such a simple self-similar scaling law the new 
scale laws to be constructed come indeed under the principle of relativity 
extended to scale transformations of the resolutions $\epsilon$, as we have 
stressed in the introduction. The structure of the corresponding 
group of scale transformations can be established from the fact that the 
involved quantities transform, under a scale transformation 
$\epsilon \rightarrow \epsilon '$, as
\be
\ln \frac{{\cal L} (\epsilon ')}{{\cal L}_0} = 
\ln \frac{{\cal L} (\epsilon)}{{\cal L}_0} +  
\delta (\epsilon)\ln \frac{\epsilon}{\epsilon '} \; ,
\label{eq.8}
\ee
\be
\delta(\epsilon ') = \delta(\epsilon).
\label{eq.9}
\ee

These transformations have exactly the structure of the Galileo group, as 
confirmed by the dilation composition law, 
$\epsilon \rightarrow \epsilon ' \rightarrow \epsilon ''$, which writes
\be
\ln {\epsilon ''\over \epsilon} = \ln {\epsilon '\over \epsilon} + 
\ln {\epsilon ''\over \epsilon '} \; . 
\label{eq.10}
\ee

It is worth noting that Eq.~(\ref{eq.5}) gives also a transition from a 
fractal to a non-fractal behavior beyond some transition scale 
$\lambda$.

\subsection{Transition from non-differentiability (fractal scales) to  
differentiability (classical scales)}
\label{ss:fracsp}

The first consequence of the 
giving up of the coordinate differentiability is the differential 
proper time symmetry breaking. Relativistic motion involves a four-dimensional 
space-time with the proper time $s$ as a curvilinear parameter. 
Strictly, the non-differentiability of the coordinates implies that the 
four-velocity
\be
V^\mu = {dX^\mu \over ds}= \lim_{ds \rightarrow 0} \frac{X^\mu (s+ds) - 
X^\mu (s)}{ds}
\label{eq.15}
\ee
is undefined. This means that, when $ds$ tends to zero,  either the ratio 
$dX^\mu /ds$ tends to infinity, or it fluctuates without reaching any limit. 
However, as recalled in the introduction, continuity and non-differentiability 
imply an explicit scale dependence of the various physical quantities, 
and therefore of the velocity, $V^\mu$. We thus replace the differential, 
$ds$, by 
a scale variable, $\delta s$, and consider $V^\mu (s,\delta s)$ as an explicit 
fractal function of this variable. For a constant fractal 
dimension $D$, the resolution in $X^\mu$, $\epsilon_\mu(\delta s)$, 
corresponds to the resolution in $s$, $\delta s$, according to
\be
\epsilon_\mu \approx \delta s ^{1/D}.
\label{eq.16}
\ee

The advantage of this method is that, for any given value of the resolution, 
$\delta s$, differentiability in $s$ is recovered, which allows to use 
the differential calculus, even when dealing with non-differentiability. \\

The scale dependence of the velocity suggests that we complete the 
standard equations of physics by new differential equations of scale. We 
therefore apply to the velocity and to the differential element, now 
interpreted as a resolution, the reasoning applied to the fractal function 
$\cal L$ in Sec.~\ref{ss:galsr}. Writing the simplest possible equation for 
the variation of the velocity $V^\mu (s, ds)$ in terms of the new scale 
variable 
$ds$, as a first order renormalization group-like differential equation 
of the form of Eq.~(\ref{eq.3}); then, Taylor-expanding it, as in 
Eq.~(\ref{eq.4}), using the fact that $V^\mu <1$ (in motion-relativistic units 
$c=1$), we obtain the solution
\be
V^\mu = v^\mu + w^\mu = v^\mu \; \left[1 + \zeta^\mu  \left(\frac{\tau}{ds}
\right)^{1-1/D}\right],
\label{eq.17}
\ee
where we have set $b=1/D-1$. Here, $v^\mu$ is the ``classical part'' of the 
velocity, 
$\overline{{\cal C} \ell}\langle V^\mu \rangle$, (see below the definition of 
the classical 
part operator $\overline{{\cal C} \ell}$), $w^\mu$ is the explicitly 
scale-dependent ``fractal part'' and $\tau$ and $\zeta^\mu$ are chosen such 
that $\overline{{\cal C} \ell}\langle\zeta^\mu\rangle = 0$ and 
$\overline{{\cal C} \ell}\langle(\zeta^\mu)^2\rangle = 1$. 
We recognize here the combination of a typical fractal behavior, with a 
fractal dimension $D$, and of a breaking of the scale symmetry at the scale 
transition $\tau$, which is 
identified with the Compton scale of the system ($\tau=\hbar/mc$), 
since $V^\mu \approx v^\mu$, when $ds \gg \tau$ (classical behavior), and 
$V^\mu \approx w^\mu$, when $ds \ll \tau$ (fractal behavior). Recalling that 
$D = 2$ plays the role of a critical dimension, we stress that, 
in the asymptotic scaling domain, $w^\mu \propto (ds/\tau)^{-1/2}$ 
\cite{LN93}, in agreement with Ref. \cite{FH65}, which allows 
to identify the fractal domain with the 
quantum one. In the present paper, only these simplest scale laws with a 
fractal dimension $D=2$ are considered. \\

The above description strictly applies to an individual fractal trajectory. 
Now, one of the geometric consequences of the non-differentiability and of 
the subsequent fractal character of space-time itself (not only of the 
trajectories) is that there is an infinity of fractal geodesics relating any 
couple of points of this fractal space-time. It has therefore been suggested 
\cite{LN89} that the description of a quantum mechanical particle, including 
its property of wave-particle duality, could be reduced to the geometric 
properties of the set of fractal geodesics that corresponds to a given state 
of this ``particle''. In such an interpretation, we do not have to endow 
the ``particle'' with internal properties such as mass, spin or charge, 
since the ``particle'' is not identified with a point mass which would 
follow the geodesics, but its internal properties can simply be 
defined as geometric properties of the fractal geodesics themselves. As a 
consequence, any measurement is interpreted as a sorting out of the 
geodesics of which the properties correspond to the resolution scale of 
the measuring device (as an example, if the ``particle" has been observed 
at a given position with a given resolution, this means that the geodesics 
which pass through this domain have been selected) \cite{LN93,LN89}. \\

The transition scale appearing in Eq.~(\ref{eq.17}) yields two 
distinct behaviors of the system (particle) depending on the resolution at 
which it is considered. 
Equation (\ref{eq.17}) multiplied by $ds$ gives the 
elementary displacement, $dX^\mu$, of the system as a sum of two terms
\be
dX^\mu = dx^\mu + d\xi^\mu,
\label{eq.18}
\ee
$d\xi^\mu$ representing the ``fractal part'' and $dx^\mu$ the ``classical 
part'', each term being defined as
\be
dx^\mu = v^\mu \; ds,
\label{eq.19}
\ee
\be
d\xi^\mu=a^\mu \sqrt{2 \cal{D}} (ds^{2})^{1/2D},
\label{eq.20}
\ee
which becomes, for $D=2$
\be
d\xi^\mu=a^\mu \sqrt{2 \cal{D}} 
ds^{1/2},
\label{eq.20bis}
\ee
with $2{\cal D}=\tau$. 
We note, from Eqs.~(\ref{eq.18}) to (\ref{eq.20bis}), that $dx^\mu$ scales 
as $ds$, while $d\xi^\mu$ scales as $ds^{1\over 2}$. Therefore,
the behavior of the system is dominated by the $d\xi^\mu$ term in the 
non-differentiable ``fractal'' domain (below the transition scale), and by 
the $dx^\mu$ one in the differentiable ``classical'' domain (above the 
transition scale). \\

Now, the Klein-Gordon and Dirac equations give results 
applying to measurements performed on quantum objects, but realised with 
classical devices, in the differentiable ``classical'' domain. The 
microphysical scale at which the physical systems under study are considered 
induces the sorting out of a bundle of geodesics, corresponding to the scale 
of the systems, while the measurement process implies a smoothing 
out of the geodesic bundle coupled to a 
transition from the non-differentiable ``fractal'' to the differentiable 
``classical'' domain. We therefore define an operator 
$\overline{{\cal C} \ell}\langle\quad\rangle$, 
which we apply to the fractal variables or functions each time we 
are drawn to the ``classical'' domain where the $ds$ behavior dominates. The 
effect of $\overline{{\cal C} \ell}$ is to extract, from the fractal 
variables or functions to which it is applied the ``classical part'', i.e., 
the part scaling as $ds$.

\subsection{Differential proper time symmetry breaking}
\label{ss:difftsb}

Another consequence of the non-differentiable nature of space-time is 
the breaking of the local differential proper time reflection invariance. The 
derivative with respect to the proper time $s$ of a differentiable function 
$f$ can be written twofold
\be
\frac{df}{ds} = \lim_{ds \rightarrow 0}\frac{f(s+ds) - f(s)}{ds} = 
\lim_{ds \rightarrow 0}\frac{f(s) - f(s-ds)}{ds} \; .
\label{eq.21}
\ee

The two definitions are equivalent in the differentiable case. One passes 
from one to the other by the transformation $ds \leftrightarrow -ds$ (local 
differential proper time reflection invariance), which is therefore an 
implicit discrete symmetry of differentiable physics. In the 
non-differentiable situation, both definitions fail, since the limits are no 
longer defined. In the new framework of scale relativity, the physics is 
related to the behavior of the function during the ``zoom'' operation on 
the proper time resolution 
$\delta s$, identified with the differential element $ds$. Two functions 
$f'_+$ and $f'_-$ are therefore defined as explicit functions of $s$ and $ds$
\be
f'_+(s,ds) = \frac{f(s+ds,ds)-f(s,ds)}{ds} \; ,
\label{eq.22}
\ee
\be
f'_-(s,ds) = \frac{f(s,ds)-f(s-ds,ds)}{ds} \; .
\label{eq.23}
\ee

When applied to the space-time coordinates, these definitions yield, in the 
non-differentiable domain, two four-velocities of which the components are 
fractal functions of the resolution, $V_+^\mu[x^\mu(s),s,ds]$ and 
$V_-^\mu[x^\mu(s),s,ds]$. In order to go back to the 
``classical'' domain and to derive the ``classical'' velocities appearing in 
Eq.~(\ref{eq.19}), we smooth out each fractal geodesic in the bundles 
selected by the zooming process with balls of radius larger than $\tau$. 
This amounts to carry out a transition from the non-differentiable to the 
differentiable domain and leads to define two ``classical'' velocity fields 
now resolution independent: 
$V_+^\mu[x^\mu(s),s,ds>\tau] = \overline{{\cal C} \ell} \langle 
V_+^\mu[x^\mu(s),s,
ds] \rangle =v_+^\mu[x^\mu(s),s]$ and 
$V_-^\mu[x^\mu(s),s,ds>\tau] = \overline{{\cal C} \ell}\langle 
V_-^\mu[x^\mu(s),s,
ds] \rangle =v_-^\mu[x^\mu(s),s]$. The important 
new fact appearing here is that, after the transition, there is no longer any 
reason for these two velocities to be equal. While, in standard mechanics, 
the concept of velocity was one-valued, we must introduce, for the case of a 
non-differentiable space, two velocities instead of one, 
even when going back to the ``classical'' domain. This two-valuedness of the 
velocity vector finds its origin in the breaking of the discrete proper 
time reflection invariance symmetry ($ds \leftrightarrow -ds$). 
Therefore, if one reverses the sign of the proper time differential element, 
$v_+^\mu$ becomes $v_-^\mu$. But the choice of a reversed proper time element 
($- ds$) must be as qualified as the initial choice ($ds$) for the 
description of the laws of nature, and we have no way, at this level of the 
description, to define what is the `` past''  and what is the ``future''. The 
only solution to this problem is to consider both the forward $ (ds > 0)$ 
and backward $ (ds < 0) $ processes on the same footing. Then the information 
needed to describe the system is doubled with respect to the standard 
differentiable description. \\

A natural way of accounting for this doubling consists in using 
complex numbers and the complex product. This is the origin of the complex 
nature of the wave function of quantum mechanics, since this wave function 
can be identified with the exponential of the complex action that is 
naturally introduced in this framework \cite{LN93,LN96}. The choice of 
complex numbers to represent the two-valuedness of 
the velocity is not an arbitrary choice. The use of these numbers is the 
most natural way to generalize to two dimensions the set of real 
numbers, and obtain, in the relativistic motion case, the complex 
Klein-Gordon equation \cite{LN96}. In the same way, as we shall see in 
Secs.~\ref{s:kgeq} and \ref{s:dieq}, the use of bi-quaternionic numbers 
(i.e., complex Hamilton's quaternions) is the most natural way to generalize 
them to eight dimensions \cite{CN02} and thus obtain the bi-quaternionic 
Klein-Gordon equation, which yields the Dirac equation. \\

\section{Bi-quaternionic Klein-Gordon equation}
\label{s:kgeq}

It has long been known that the Dirac equation naturally proceeds from the 
Klein-Gordon equation when writen in a quaternionic form \cite{CL29,AC37}, 
i.e., using the quaternionic formalism, as introduced by Hamilton 
\cite{WH66}, and further developed by Conway \cite{AC37,AC45}. However, this 
remains essentially formal in the standard framework, since there is no first 
principle reason for which the probability amplitude should be quaternionic. 
In the present approach, we shall see that its bi-quaternionic nature can be 
established as a consequence of the nonderivable nature of space-time. 
Indeed, we propose in the current section to introduce naturally a 
bi-quaternionic covariant derivative operator, leading to the definition of a 
bi-quaternionic velocity and wave-function, which we use to derive the 
Klein-Gordon equation in a bi-quaternionic form. This allows us to obtain, 
in Sec.~\ref{s:dieq}, the Dirac equation as a mere consequence, since Dirac 
spinors and bi-quaternions are actually two equivalent representations 
of an electron. \\

\subsection{Further symmetry breakings and bi-quaternionic covariant 
derivative operator}
\label{ss:bqdop}

Most of the approach described in Sec.~\ref{s:schro} remains 
applicable. However, the main new features obtained in the case we now 
study proceed from a deeper description of the scale formalism, 
considering the more general case when the peculiar choice of 
an identification of the  differentials and the resolution variables 
(remember the choice explicited in Sec.~\ref{ss:fracsp}, where we have set 
$ds=\delta s$) is given up, implying the subsequent breaking of the 
symmetries: 
\begin{center}
$ds\leftrightarrow -ds \qquad 
dx^{\mu}\leftrightarrow -dx^{\mu} \qquad 
x^{\mu}\leftrightarrow -x^{\mu}$ \\
\end{center}

 \begin{figure*}
   \centering
\includegraphics{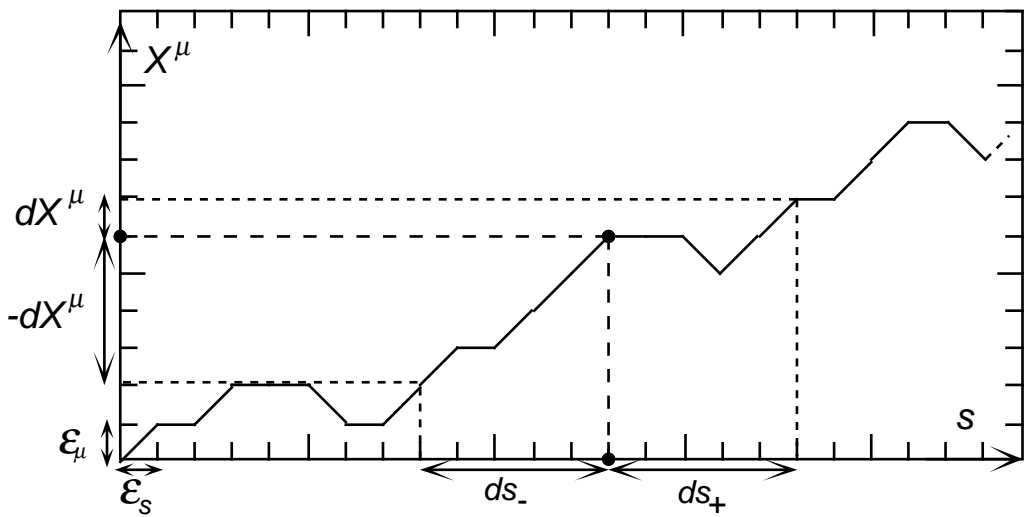}

   \caption{Fractal function $X^{\mu}(s, \epsilon_{\mu}, \epsilon_{s})$}
              \label{fig.1}
    \end{figure*}

In the scaling domain, the four space-time coordinates 
$X^\mu(s, \epsilon_{\mu}, \epsilon_{s})$ are now four fractal functions of the 
proper time $s$ and of the resolutions $\epsilon_\mu$ for the coordinates and 
$\epsilon_s$ for the proper time. We consider the case when, for an 
elementary displacement $dX^\mu$ corresponding to a shift $ds$ in the 
curvilinear parameter, the resolutions verify $\epsilon_{\mu}<dX^{\mu}$ and 
$\epsilon_{s}<ds$, which implies that, at 
a given $X^\mu$, a forward shift $ds$ of $s$ yields a displacement $dX^\mu$ 
of $X^\mu$ and a backward shift $-ds$ produces a displacement $-dX^\mu$, of 
which the amplitudes are not necessarily equal (see Fig.~\ref{fig.1}). \\

We can therefore apply, to these two different elementary displacements, 
the canonical decomposition
\begin{eqnarray}
dX^{\mu}=dx^{\mu}+d\xi^{\mu}, \qquad {\rm with} \qquad 
\overline{{\cal C} \ell} \langle dX^\mu \rangle =dx^{\mu} = v^{\mu}_{+\atop{\mu}} \; ds,
\nonumber \\
d\xi^{\mu}=a^{\mu}_+ \sqrt{2 \cal{D}} (ds^{2})^{\frac{1}{2D}}, \qquad 
\overline{{\cal C} \ell} \langle a^{\mu}_+ \rangle =0, \qquad \overline{{\cal C} \ell}\langle(a^{\mu}_+)^2\rangle=1,
\label{eq.60}
\end{eqnarray}
and
\begin{eqnarray}
-dX^{\mu}=\delta x^{\mu}+\delta \xi^{\mu}, \qquad {\rm with} \qquad 
\overline{{\cal C} \ell} \langle -dX^\mu \rangle =\delta x^{\mu} = v^{\mu}_{-\atop{\mu}} \; ds,
\nonumber \\
\delta \xi^{\mu}=a^{\mu}_- \sqrt{2 \cal{D}} (ds^{2})^{\frac{1}{2D}}, \qquad 
\overline{{\cal C} \ell }\langle a^{\mu}_- \rangle =0, \qquad \overline{{\cal C} \ell} \langle (a^{\mu}_-)^2 \rangle =1.
\label{eq.63}
\end{eqnarray}

In the differentiable case, $dX^\mu=-(-dX^\mu)$, and therefore 
$v^{\mu}_{+\atop{\mu}}=-v^{\mu}_{-\atop{\mu}}$. This is no longer the 
case in the non-differentiable case, where the local symmetry 
$dx^{\mu}\leftrightarrow -dx^{\mu}$ is broken. \\

Furthermore, we must also consider the breaking of the symmetry 
$ds\leftrightarrow -ds$ proceeding from the twofold definition of the 
derivative with respect to the curvilinear parameter $s$. Applied to $X^\mu$, 
considering an elementary displacement $dX^\mu$, the classical part 
extraction process gives two classical forward and backward 
derivatives $d/ds_+$ and $d/ds_-$, which yield in turn two classical 
velocities, which we denote $v^\mu_{{\pm \atop{s}} {+\atop{\mu}}}$. 
Considering now the same extraction process, applied to an elementary 
displacement $-dX^\mu$, the classical forward and backward derivatives 
$d/ds_+$ and $d/ds_-$ allow again to define two classical velocities, 
denoted $v^\mu_{{\pm \atop{s}} {-\atop{\mu}}}$. We summarize this result as 
\be
v^{\mu}_{{\pm \atop{s}} {+\atop{\mu}}}={dx^\mu\over {ds_{\pm}}}, \qquad 
v^{\mu}_{{\pm \atop{s}} {-\atop{\mu}}}={\delta x^\mu\over {ds_{\pm}}} \; .
\label{eq.64}
\ee

Contrary to what happens in the differentiable case, the total derivative 
with respect to the proper time of a fractal function $f(x(s),s)$ with 
integer fractal dimension contains finite terms up to higher order \cite{AE05} 
\be
{df\over {ds}}= {\partial f \over {\partial s}} + {\partial f \over 
{\partial x^\mu}}{dX^\mu \over {ds}} + {1\over 2}{\partial^2 f \over {\partial 
x^\mu \partial x^\nu}}{dX^\mu dX^\nu \over {ds}} + {1\over 6}{\partial^3 
f \over {\partial x^\lambda \partial x^\mu \partial x^\nu}}{dX^\lambda dX^\mu
dX^\nu \over {ds}} + ...
\label{eq.29}
\ee

In our case, a finite contribution only proceeds from terms of $D$-order, 
while lesser-order terms yield an infinite contribution and higher-order ones 
are negligible. Therefore, in the special case of a fractal dimension $D=2$, 
the total derivative writes
\be
{df\over {ds}} = \frac{\partial f}{\partial s} + \partial_\mu f . {dX^\mu
\over {ds}} + \frac{1}{2} \frac{\partial ^2 f}{\partial x^\mu \partial x^\nu} 
{dX^\mu dX^\nu \over {ds}} \; .
\label{eq.30}
\ee

Usually the term $dX^\mu dX^\nu /ds$ is infinitesimal, but here its classical 
part reduces to $\overline{{\cal C} \ell} \langle d \xi^\mu\; d\xi^\nu 
\rangle /ds$. \\

We can, at this stage, define several total derivatives with respect 
to $s$ of a fractal function $f$. We write them - using a compact 
straightforward notation with summation over repeated indices - after 
substituting, in the four-dimensional 
analog of Eq.~(\ref{eq.30}), the expressions for the derivatives of the 
$X^\mu$ obtained when using Eqs.~(\ref{eq.60}) and (\ref{eq.63}) with the 
different expressions of Eq.~(\ref{eq.64}) for the classical velocities
\be
{df\over {ds}}{}_{{\pm \atop{s}} {\pm \atop{x}} {\pm \atop{y}} 
{\pm \atop{z}} {\pm \atop{t}}}= {\partial f\over {\partial s}} + 
(v^{\mu}_{{\pm \atop{s}} {\pm \atop{\mu}}} + w^{\mu}_{{\pm \atop{s}} 
{\pm \atop{\mu}}}){\partial f\over {\partial X^\mu}} + a^\mu_\pm a^\nu_\pm 
{\cal D}{\partial^2 f\over {dX^\mu dX^\nu}} \; ,
\label{eq.65}
\ee
\be
{\rm with} \qquad 
w^\mu=a^\mu {\sqrt {2 {\cal D}}} (ds^2)^{{1\over {2D}}-{1\over 2}}.
\label{eq.66}
\ee

Now, when we apply the classical part operator to Eq.~(\ref{eq.65}), using 
Eq.~(\ref{eq.66}) and the 
properties of the $a^\mu_{\pm}$ 's as stated in Eqs.~(\ref{eq.60}) and 
(\ref{eq.63}), the $w^\mu$ 's disappear at the first order, but, at the second 
order, for the fractal dimension $D=2$, the fractal behavior of 
Eq.~(\ref{eq.20}) writes
\be
\overline{{\cal C} \ell} \langle w^{\mu}_{{\pm \atop{s}} {\pm \atop{\mu}}} w^{\nu}_
{{\pm \atop{s}} {\pm \atop{\nu}}} \rangle =\mp 2 {\cal D} \eta^{\mu \nu} ds ,
\label{eq.67b}
\ee
the $\mp$ sign in the right-hand side being the inverse of the s-sign in 
the left-hand side. \\

Equation~(\ref{eq.67b}) follows from the fact that the $d\xi^\mu$'s are of 
null classical part and mutually independent. The Minkowski metric component, 
$\eta^{\mu \nu}$ implies indeed that the classical part of every crossed 
product $w^{\mu}_{{\pm \atop{s}} {\pm \atop{\mu}}} w^{\nu}_
{{\pm \atop{s}} {\pm \atop{\nu}}}$, with $\mu \neq \nu$, is null. This is 
due to the fact that, even if each term in the product scales as $ds^{1/2}$, 
each behaves as an independent fractal fluctuation around its own classical 
part. Therefore, when we proceed to the smoothing out of the geodesic bundle 
during the transition from the fractal to the classical domain, we apply a 
process which is mathematically (not physically) equivalent to a stochastic 
``Wiener'' process, and also more general, since we do not need any Gaussian 
distribution assumption. Thus, we can apply to the classical part of the 
$w$ crossed product the property of the product of two independent 
stochastic variables: i.e., the classical part of the product is the product 
of the classical parts, namely here zero. \\

Thanks to Eq.~(\ref{eq.67b}), the last term of the classical part of 
Eq.~(\ref{eq.65}) amounts to a Dalambertian, and we can write  
\be
{df\over {ds}}{}_{{\pm \atop{s}} {\pm \atop{x}} {\pm \atop{y}} 
{\pm \atop{z}} {\pm \atop{t}}}= \left({\partial \over {\partial s}} + 
v^{\mu}_{{\pm \atop{s}} {\pm \atop{\mu}}} {\partial _\mu} 
\mp {\cal D}{\partial^\mu \partial_\mu}\right) f \; ,
\label{eq.67c}
\ee
where the $\mp$ sign in the right-hand side is still the inverse of the 
s-sign. When we apply these derivatives to the position vector $X^\mu$, we 
obtain, as expected,
\be
{dX^{\mu}\over {ds}}{}_{{\pm \atop{s}} {\pm \atop{\mu}}} = v^{\mu}_{{\pm 
\atop{s}} {\pm \atop{\mu}}} .
\label{eq.67}
\ee

We consider now the four fractal functions $-X^\mu(s, \epsilon_\mu, 
\epsilon_s)$. At this description level, there is no reason for 
$(-X^{\mu})(s, \epsilon_\mu, \epsilon_s)$ to be everywhere equal to 
$-(X^{\mu})(s, \epsilon_\mu, \epsilon_s)$, owing to a local 
breaking of the P (for $\mu = x,y,z$) and T (for $\mu = t$) symmetries. 
Furthermore, as we have stressed above for $X^\mu$, 
at a given $-X^\mu$, a forward shift $ds$ of $s$ yields a displacement 
$d(-X^\mu)$ of $-X^\mu$ and a backward shift $-ds$ produces a displacement 
$-d(-X^\mu)$, with no necessarily equal amplitudes. We can therefore apply, 
to these two different elementary displacements, a decomposition similar to 
the one described in Eqs.~(\ref{eq.60}) and (\ref{eq.63}), i.e.,
\begin{eqnarray}
d(-X^{\mu})={\tilde d} x^{\mu}+{\tilde d}\xi^{\mu}, \qquad {\rm with} \qquad 
\overline{{\cal C} \ell} \langle d(-X^\mu) \rangle ={\tilde d}x^{\mu} = 
{\tilde v}^{\mu}_{+\atop{\mu}} \; ds, \nonumber \\
{\tilde d}\xi^{\mu}={\tilde a}^{\mu}_+ \sqrt{2 \cal{D}} 
(ds^{2})^{\frac{1}{2D}}, \qquad \overline{{\cal C} \ell} \langle {\tilde a}^{\mu}_+ \rangle 
=0, \qquad \overline{{\cal C} \ell} \langle ({\tilde a}^{\mu}_+)^2 \rangle =1,
\label{eq.70}
\end{eqnarray}
and
\begin{eqnarray}
-d(-X^{\mu})={\tilde \delta} x^{\mu}+{\tilde \delta} \xi^{\mu}, 
\qquad {\rm with} \qquad 
\overline{{\cal C} \ell} \langle -d(-X^\mu) \rangle ={\tilde \delta} x^{\mu} = 
{\tilde v}^{\mu}_{-\atop{\mu}} \; ds, \nonumber \\
{\tilde \delta} \xi^{\mu}={\tilde a}^{\mu}_- \sqrt{2 \cal{D}} 
(ds^{2})^{\frac{1}{2D}}, \qquad \overline{{\cal C} \ell} \langle {\tilde a}^{\mu}_- \rangle 
=0, \qquad \overline{{\cal C} \ell} \langle ({\tilde a}^{\mu}_-)^2 \rangle =1,
\label{eq.73}
\end{eqnarray}

Then, we jump to the classical parts and we are once more confronted to the 
breaking of the $ds\leftrightarrow -ds$ symmetry. Applied to $-X^\mu$, 
considering an elementary displacement $d(-X^\mu)$, the classical forward 
and backward derivatives $d/ds_+$ and $d/ds_-$ give again two ``classical'' 
velocities, which we denote ${\tilde v}^\mu_{{\pm \atop{s}} {+\atop{\mu}}}$. 
After another classical part extraction, considering now an elementary 
displacement $-d(-X^\mu)$, the classical forward and backward derivatives 
$d/ds_+$ and $d/ds_-$ yield two other ``classical'' velocities, denoted 
${\tilde v}^\mu_{{\pm \atop{s}} {-\atop{\mu}}}$. We write in short
\be
{\tilde v}^{\mu}_{{\pm \atop{s}} {+\atop{\mu}}}={{\tilde d}x^\mu\over 
ds_{\pm}}, \qquad {\tilde v}^{\mu}_{{\pm \atop{s}} {-\atop{\mu}}}=
{{\tilde \delta} x^\mu\over {ds_{\pm}}} \; .
\label{eq.74}
\ee

We therefore obtain new different total derivatives with respect to $s$ 
of a fractal function $f$, which we write
\be
{{\tilde d}f\over {ds}}{}_{{\pm \atop{s}} {\pm \atop{x}} {\pm \atop{y}} 
{\pm \atop{z}} {\pm \atop{t}}}= {\partial f\over {\partial s}} + 
({\tilde v}^{\mu}_{{\pm \atop{s}} {\pm \atop{\mu}}} + {\tilde w}^{\mu}_{{\pm 
\atop{s}} {\pm \atop{\mu}}}){\partial f\over {\partial X^\mu}} + 
{\tilde a}^\mu_\pm {\tilde a}^\nu_\pm {\cal D}{\partial^2 f\over 
{dX^\mu dX^\nu}} \; ,
\label{eq.75}
\ee
\be
{\rm with} \qquad {\tilde w}^\mu={\tilde a}^\mu {\sqrt {2 {\cal D}}} 
(ds^2)^{{1\over {2D}}-{1\over 2}}.
\label{eq.76}
\ee

The classical part operator applied to Eq.~(\ref{eq.75}) yields the classical 
total derivatives
\be
{{\tilde d}f\over {ds}}{}_{{\pm \atop{s}} {\pm \atop{x}} {\pm \atop{y}} 
{\pm \atop{z}} {\pm \atop{t}}}= \left({\partial \over {\partial s}} + 
{\tilde v}^{\mu}_{{\pm \atop{s}} {\pm \atop{\mu}}} {\partial _\mu} 
\mp {\cal D}{\partial^\mu \partial_\mu}\right) f \; .
\label{eq.77b}
\ee

When we apply these derivatives to the position vector $X^\mu$, we obtain, as 
expected again,
\be
{{\tilde d}X^{\mu}\over {ds}}{}_{{\pm \atop{s}} {\pm \atop{\mu}}} = 
{\tilde v}^{\mu}_{{\pm \atop{s}} {\pm \atop{\mu}}}.
\label{eq.77}
\ee

If we consider the simplest peculiar case when the breaking of the symmetry 
$dx^{\mu}\leftrightarrow -dx^{\mu}$ is isotropic as regards the four 
space-time coordinates (i.e., the signs corresponding to the four $\mu$ 
indices are chosen equal), we are left with eight non-degenerate 
components - four $v^{\mu}_{{\pm \atop{s}} {\pm \atop{\mu}}}$ and  
four ${\tilde v}^{\mu}_{{\pm \atop{s}} {\pm \atop{\mu}}}$ - which can be 
used to define a bi-quaternionic four-velocity,
\begin{eqnarray}
{\cal V}^\mu=&&{1\over 2}(v^\mu_{++} + {\tilde v}^\mu_{--})-{i\over 2}
(v^\mu_{++} - {\tilde v}^\mu_{--}) +\left[{1\over 2}(v^\mu_{+-} + 
v^\mu_{-+})-{i\over 2}(v^\mu_{+-} - {\tilde v}^\mu_{++})\right] e_1 
\nonumber \\
+&& \left[{1\over 2}(v^\mu_{--} + {\tilde v}^\mu_{+-})-{i\over 2}(v^\mu_{--} - 
{\tilde v}^\mu_{-+})\right] e_2 + \left[{1\over 2}(v^\mu_{-+} + 
{\tilde v}^\mu_{++})-{i\over 2}({\tilde v}^\mu_{-+} + 
{\tilde v}^\mu_{+-})\right] e_3 .
\label{eq.78}
\end{eqnarray}
with $e_i (i=1,2,3)$ denoting Hamilton's imaginary units satisfying the 
associative but non-commutative algebra 
\be
e_ie_j= -\delta_{ij} + \sum_{k=1}^3\epsilon_{ijk}e_k, \qquad i,j=1,2,3
\label{eq.106}
\ee
where $\epsilon_{ijk}$ is the usual completely antisymmetric three-index 
tensor with $\epsilon_{123}=1$. \\

The freedom in the choice of the actual expression for ${\cal V}^\mu$ will 
be discussed later. It is constrained fy the following requirements: at the 
limit when $\epsilon_\mu \rightarrow dX^\mu$ and $\epsilon_s 
\rightarrow ds$, every $e_i$-term in Eq.~(\ref{eq.78}) goes to zero, 
and, as ${\tilde v}^\mu_{--} = v^\mu_{-+}$ in this limit, one 
recovers the complex velocity of the scale-relativistic Schr\"odinger 
equation \cite{LN93}, 
${\cal V}^\mu=[v^\mu_{++}+v^\mu_{-+}-i(v^\mu_{++}-v^\mu_{-+})]/2$; at the 
classical limit, every term in this equation vanishes, save the real term, 
and the velocity becomes classical in its usual meaning, i.e., real: 
${\cal V}^\mu=v^\mu_{++}$. \\

The bi-quaternionic velocity thus defined corresponds to a bi-quaternionic 
derivative operator $\ddfr /ds$, similarly defined, and yielding, 
when applied to the position vector $X^\mu$, the corresponding velocity. For 
instance, the derivative operator attached to the velocity in 
Eq.~(\ref{eq.78}) writes
\begin{eqnarray}
{\ddfr \over {ds}}={1\over 2}\left({d\over {ds}}{}_{++}+{{\tilde d}\over 
{ds}}{}_{--}\right) -{i\over 2}\left({d\over {ds}}{}_{++} - {{\tilde d}\over 
{ds}}{}_{--}\right) + \left[{1\over 2}\left({d\over {ds}}{}_{+-} + {d\over 
{ds}}{}_{-+}\right) - {i\over 2}\left({d\over {ds}}{}_{+-} - {{\tilde d}\over 
{ds}}{}_{++}\right)\right] e_1 \nonumber \\
+ \left[{1\over 2}\left({d\over {ds}}{}_{--} + {{\tilde d}\over 
{ds}}{}_{+-}\right) - {i\over 2}\left({d\over {ds}}{}_{--} - {{\tilde d}\over 
{ds}}{}_{-+}\right)\right] e_2 + \left[{1\over 2}\left({d\over {ds}}{}_{-+}+
{{\tilde d}\over {ds}}{}_{++}\right) -{i\over 2}\left({{\tilde d}\over 
{ds}}{}_{-+} + {{\tilde d}\over {ds}}{}_{+-}\right)\right] e_3 . \nonumber \\
\label{eq.79}
\end{eqnarray}

Substituting Eqs.~(\ref{eq.67c}) and (\ref{eq.77b}) into 
Eq.~(\ref{eq.79}), we obtain the expression for the bi-quaternionic 
proper-time derivative operator 
\be
{\ddfr \over {ds}}= [1+e_1+e_2+(1-i)e_3]{\partial \over {\partial s}} + 
{\cal V}^\mu \partial_\mu + i{\cal D} \partial ^\mu \partial _\mu ,
\label{eq.80}
\ee
the $+$ sign in front of the Dalambertian proceeding from the choice of the 
metric signature $(+,-,-,-) $. We keep here, for generality, the $\partial 
/ \partial s$ term, stressing that it is actually of no use, since the 
various physical functions are not explicitly depending on $s$. 
It is easy to check that this operator, applied to the position vector 
$X^\mu$, gives back the bi-quaternionic velocity ${\cal V}^\mu$ of 
Eq.~(\ref{eq.78}). \\

It is worth noting that the expression we have written for ${\cal V}^\mu$ 
in Eq.~(\ref{eq.78}) is one among the various choices we could have 
retained to define the bi-quaternionic velocity. The main constraint limiting 
this choice is the recovery of the complex and real velocities at the 
non-relativistic motion and classical limits. We also choose ${\cal V}^{\mu}$ 
such as to obtain the third term in the right-hand side of Eq.~(\ref{eq.80}) 
under the form of a purely imaginary Dalembertian, which allows to recover 
an integrable equation of motion. To any bi-quaternionic velocity satisfying 
both prescriptions corresponds a bi-quaternionic derivative operator 
$\ddfr /ds$, similarly defined, and yielding 
this velocity when applied to the position vector $X^\mu$. But, 
whatever the definition retained, the derivative operator keeps 
the same form in terms of the bi-quaternionic velocity ${\cal V}^\mu$, as 
given by Eq.~(\ref{eq.89}). 
Therefore, the different choices allowed for its definition merely correspond 
to different mathematical representations leading to the same physical 
result.

\subsection{Bi-quaternionic wave-function}
\label{ss:bqsap}

Since ${\cal V}^\mu$ is bi-quaternionic, the lagrange function is also 
bi-quaternionic and, therefore, the same is true of the action. \\

A generalized equivalence principle, as well as a strong covariance 
principle, allows us to write the equation of motion under a free-motion 
form, i.e., the differential geodesic equation
\be
{\ddfr {\cal V}_{\mu} \over {ds}}=0,
\label{eq.84b}
\ee
where ${\cal V}_{\mu}$ is the bi-quaternionic four-velocity, e.g., the 
covariant counterpart of ${\cal V}^\mu$ defined in Eq.~(\ref{eq.78}). \\

The elementary variation of the action, considered as a functional of the 
coordinates, keeps the usual form
\be
\delta {\cal S}= - mc \; {\cal V}_{\mu} \; \delta x^{\mu}.
\label{eq.85b}
\ee

We thus obtain the bi-quaternionic four-momentum, as
\be
{\cal P}_{\mu}=mc{\cal V}_{\mu}= -\partial_{\mu}{\cal S}.
\label{eq.86b}
\ee

We are now able to introduce the wave function. We define it as 
a re-expression of the bi-quaternionic action by
\be
\psi^{-1} \partial_{\mu} \psi = {i\over {c S_0}} \partial_{\mu} {\cal S},
\label{eq.87b}
\ee
using, in the left-hand side, the quaternionic product 
derived from Eq.~(\ref{eq.86b}), as
\be
{\cal V}_{\mu}=i{S_0\over m} \psi^{-1} \partial_{\mu} \psi.
\label{eq.88b}
\ee 

This is the bi-quaternionic generalization of the definition used in the 
Schr\"odinger case: $\psi=e^{iS/S_0}$. It is worth stressing here that we 
could choose, for the definition of the 
wave function in Eq.~(\ref{eq.87b}), a commutated expression in the left-hand 
side, i.e., $(\partial_{\mu} \psi) \psi^{-1}$ instead of $\psi^{-1} 
\partial_{\mu} \psi$. But with this reversed choice, 
owing to the non-commutativity of the quaternionic product, we 
could not obtain the motion equation as a vanishing four-gradient, as in 
Eq.~(\ref{eq.95}). Therefore, we retain the above simplest choice, 
as yielding an equation which can be integrated. This non-commutativity 
induced property might be considered as a key for a future understanding of 
the parity violation, which will not be developed here.

\subsection{Free-particle Klein-Gordon equation}
\label{ss:fpkg}

As, in what follows, we 
only consider $s$-stationary functions, i.e., functions which do not 
explicitly depend on the proper time $s$, the derivative operator reduces 
to
\be
{\ddfr \over {ds}}={\cal V}^\nu \partial_\nu + i{\cal D} \partial ^\nu 
\partial _\nu.
\label{eq.89}
\ee

Now this expression is independent of the peculiar 
choice retained for the bi-quaternionic form of the four-velocity, as the 
representation dependent term in the right-hand side of Eq.~(\ref{eq.80}) 
has vanished. \\

The equation of motion, Eq.~(\ref{eq.84b}), thus writes 
\be
\left ({\cal V}^\nu \partial_\nu + i{\cal D} \partial ^\nu 
\partial _\nu \right ) {\cal V}_{\mu} = 0.
\label{eq.90}
\ee

We replace ${\cal V}_{\mu}$, respectively ${\cal V}^{\nu}$, by their 
expressions given in Eq.~(\ref{eq.88b}) and obtain 
\be
i{S_0\over m} \left ( i{S_0\over m} \psi^{-1} \partial^{\nu} \psi 
\partial_\nu + i{\cal D} \partial ^\nu \partial _\nu \right ) \left ( 
\psi^{-1} \partial_{\mu} \psi \right ) = 0.
\label{eq.91}
\ee

The particular choice ${\cal S}_0= 2 m {\cal D}$ \footnote {At first sight, 
the meaning of the choice ${\cal S}_{0}=2 m 
{\cal D}$ is not a necessary condition from the 
viewpoint of physics, but merely a simplifying choice as regards the equation 
form. Indeed, it is only under this particular choice that the 
fundamental equation of dynamics can be integrated. However, if we do not 
make this choice, the $\psi$ function is a solution of a third order, non 
linear, complicated equation such that no precise physical meaning can be 
given to it. We therefore claim that our choice ${\cal S}_{0}=2 m {\cal D}$ 
has a profound physical significance, since, in our construction, the 
meaning of $\psi$ is directly related to the fact that it is a solution of the 
Klein-Gordon and Dirac equations obtained below.} allows 
us to simplify this equation and we get
\be
\psi^{-1} \partial^{\nu} \psi \; \partial_\nu (\psi^{-1}  
\partial_{\mu} \psi) + {1\over 2} \partial ^\nu \partial _\nu (\psi^{-1} 
\partial_{\mu} \psi) = 0.
\label{eq.92}
\ee

The definition of the inverse of a quaternion
\be
\psi \psi^{-1} =  \psi^{-1} \psi = 1,
\label{eq.93}
\ee
implies that $\psi$ and $\psi^{-1}$ commute. But this is not necessarily 
the case for $\psi$ and $\partial _{\mu} \psi^{-1}$ nor for $\psi^{-1}$ and 
$\partial _{\mu} \psi$ and their contravariant counterparts 
. However, when we derive Eq.~(\ref{eq.93}) with respect to 
the coordinates, we obtain
\begin{eqnarray}
\psi \; \partial _{\mu} \psi^{-1} &=& - (\partial _{\mu} \psi) \psi^{-1} 
\nonumber \\
\psi^{-1} \partial _{\mu} \psi &=& - (\partial _{\mu} \psi^{-1}) \psi,
\label{eq.94}
\end{eqnarray}
and identical formulae for the contravariant analogues. \\

Developing Eq.~(\ref{eq.92}), using Eqs.~(\ref{eq.94}) and the property 
$\partial^{\nu}\partial_{\nu} \partial_{\mu} = \partial_{\mu}\partial^{\nu} 
\partial_{\nu}$, we obtain, after some calculations,
\be
\partial_{\mu}[(\partial^{\nu}\partial_{\nu} \psi) \psi^{-1}] = 0.
\label{eq.95}
\ee

We integrate this four-gradient as
\be
(\partial^{\nu}\partial_{\nu} \psi) \psi^{-1} + C = 0 ,
\label{eq.96}
\ee
of which we take the right product by $\psi$ to obtain 
\be
\partial^{\nu}\partial_{\nu} \psi + C \psi = 0.
\label{eq.97}
\ee

We therefore recognize the Klein-Gordon equation for a free particle with 
a mass $m$, so that $m^2c^2/{\hbar}^2=C$. \\

\section{Dirac equation}
\label{s:dieq}

We now use a long-known property of the quaternionic formalism, which allows 
to obtain the Dirac equation for a free particle as a mere square root of 
the Klein-Gordon operator (see, e.g., Refs. \cite{CL29,AC37}). \\

We first develop the Klein-Gordon equation as
\be
{1\over {c^2}}{\partial^2 \psi \over {\partial t^2}} = {\partial^2 \psi \over 
{\partial x^2}} + {\partial^2 \psi \over {\partial y^2}} + 
{\partial^2 \psi \over {\partial z^2}} - {m^2c^2\over {\hbar^2}} \psi.
\label{eq.98}
\ee

Thanks to the property of the quaternionic and complex imaginary units 
$e^2_1=e^2_2=e^2_3=i^2=-1$, we can write Eq.~(\ref{eq.98}) under the form
\be
{1\over {c^2}}{\partial^2 \psi \over {\partial t^2}} = e^2_3{\partial^2 \psi 
\over {\partial x^2}}e^2_2 + ie^2_1{\partial^2 \psi \over {\partial y^2}}i + 
e^2_3{\partial^2 \psi \over {\partial z^2}}e^2_1 + i^2{m^2c^2\over {\hbar^2}} 
e^2_3 \psi e^2_3.
\label{eq.99}
\ee

We now take advantage of the anticommutative property of the quaternionic 
units ($e_ie_j=-e_je_i$ for $i\neq j$) to add to the right-hand side of 
Eq.~(\ref{eq.99}) six vanishing couples of terms which we rearrange as
\begin{eqnarray}
{1\over c}{\partial\over {\partial t}} \left ({1\over c}{\partial \psi 
\over {\partial t}}\right ) &=& e_3 {\partial\over {\partial x}} \left ( 
e_3 {\partial \psi \over {\partial x}}e_2 + e_1 {\partial \psi \over 
{\partial y}}i + e_3 {\partial \psi \over {\partial z}}e_1 - 
i{mc\over \hbar}e_3 \psi e_3 \right )e_2 \nonumber \\
&+& e_1 {\partial\over {\partial y}} \left ( 
e_3 {\partial \psi \over {\partial x}}e_2 + e_1 {\partial \psi \over 
{\partial y}}i + e_3 {\partial \psi \over {\partial z}}e_1 - 
i{mc\over \hbar}e_3 \psi e_3 \right )i \nonumber \\
&+& e_3 {\partial\over {\partial z}} \left ( 
e_3 {\partial \psi \over {\partial x}}e_2 + e_1 {\partial \psi \over 
{\partial y}}i + e_3 {\partial \psi \over {\partial z}}e_1 - 
i{mc\over \hbar}e_3 \psi e_3 \right )e_1 \nonumber \\
&-& i{mc\over \hbar}e_3 \left ( 
e_3 {\partial \psi \over {\partial x}}e_2 + e_1 {\partial \psi \over 
{\partial y}}i + e_3 {\partial \psi \over {\partial z}}e_1 - 
i{mc\over \hbar}e_3 \psi e_3 \right )e_3.
\label{eq.100}
\end{eqnarray}

We see that Eq.~(\ref{eq.100}) is obtained by 
applying twice to the bi-quaternionic wavefunction $\psi$ the operator
\be
{1\over c}{\partial\over {\partial t}} = e_3 {\partial \over 
{\partial x}}e_2 + e_1 {\partial \over 
{\partial y}}i + e_3 {\partial \over {\partial z}}e_1 - 
i{mc\over \hbar}e_3 ( \quad ) e_3.
\label{eq.101}
\ee

The three first Conway matrices $e_3(\quad)e_2$, $e_1(\quad)i$ and 
$e_3(\quad)e_1$ \cite{JS72}, figuring in the right-hand side 
of Eq.~(\ref{eq.101}), can be written in the compact form $-\alpha^k$, 
with 
\begin{displaymath}
\alpha^k= \left (
          \begin{array}{cc}
           0 & \sigma_k \\ 
           \sigma_k & 0 
           \end{array} 
           \right ), 
\end{displaymath}
the $\sigma_k$'s being the three Pauli matrices, while the fourth Conway 
matrix
\begin{displaymath}
e_3(\quad)e_3= \left (
          \begin{array}{cccc}
           1 & \quad 0 & \; \; 0 & 0 \\ 
           0 & \quad 1 & \; \; 0 & 0 \\
           0 & \quad 0 & \; \; -1 & 0 \\
           0 & \quad 0 & \; \; 0 & -1 
           \end{array}
           \right ) 
\end{displaymath}
can be recognized as the Dirac $\beta$ matrix. We can therefore write 
Eq.~(\ref{eq.101}) as the non-covariant Dirac equation for a free fermion
\be
{1\over c}{\partial \psi \over {\partial t}} = - \alpha^k{\partial \psi \over 
{\partial x^k}} - i{mc\over \hbar}\beta \psi .
\label{eq.102}
\ee

The covariant form, in the Dirac representation, can be recovered by 
applying $ie_3(\quad)e_3$ to Eq.~(\ref{eq.102}). \\

The isomorphism which can be established between the quaternionic and  
spinorial algebrae through the multiplication rules applying to the Pauli 
spin matrices allows us to identify the wave function $\psi$ 
to a Dirac spinor. Spinors and quaternions are both a representation of 
the SL(2,C) group. See Ref. \cite{PR64} for a detailed discussion of the 
spinorial properties of bi-quaternions.

\section{Conclusion}
\label{s:conc}

The last of the three most fundamental motion equations (Schr\"odinger, 
Klein-Gordon and Dirac) which are merely postulated in standard quantum 
mechanics has been established here, in the framework of Galilean scale 
relativity, as a geodesic equation in a fractal space-time. 
The change from classical to 
quantum relativistic motion arises from successive symmetry breakings in 
the fractal geodesic picture. \\

First, the complex nature of the wave function is the 
result of the differential (proper) time symmetry breaking, which is  
the simplest effect arising from the fractal structure of space (space-time). 
At this stage, Galilean scale relativity with a 
complex wave function permits the derivation of both the Schr\"odinger 
\cite{LN93} and Klein-Gordon \cite{LN96} equations. \\

To go on with the description of the elementary properties encountered 
in the microphysical world, we have considered here further breakings of 
fundamental discrete symmetries, due to nonderivability, namely the 
differential coordinate symmetry 
($dx^{\mu} \leftrightarrow -dx^{\mu}$) breaking and the parity and 
time reversal symmetry breakings. These new breakings provide a four-complex 
component wave function (i.e., a eight component wave function), of which 
the most natural mathematical representation is in term of bi-quaternionic 
numbers \cite{CN02}. We therefore obtain the spinorial 
and the particle anti-particle nature of elementary objects which we can 
describe as Dirac spinors. Here, spin arises from the isomorphism between 
the quaternionic and spinorial representations, both of which are different 
representations of the SL(2,C) group. At this stage, the Klein-Gordon 
equation is written in a 
bi-quaternionic form which naturally yields the free Dirac equation. \\

It is worth stressing that these results only proceed from 
a restricted use of the scale-relativistic potentialities. We have, in 
the present work, limited our investigations to the induced effects of 
scale laws on the equations of motion, 
in the framework of dilation laws exhibiting a Galilean group structure, i.e., 
a fractal space-time with a constant fractal dimension $D=2$. This is only 
one of the simplest levels at 
which the scale-relativistic program can be achieved. We have also made a 
series of simplifying choices, which we have explained and justified all along 
the derivation procedure. Some of them have been dictated by physical or 
experimental considerations, but others only correspond to special cases, 
provisionally retained, so that  other possibilities will have to be explored 
in the future. \\

\end{document}